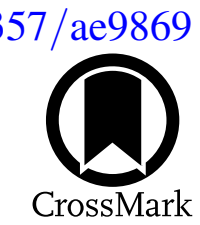

# High-$CO_2$ Climates and Observables in the Outer Habitable Zone

Daria Pidhorodetska[1,8], Edward W. Schwieterman[1,2], Thomas J. Fauchez[3,4,5], and Martin Turbet[6,7]
[1] Department of Earth and Planetary Sciences, University of California, Riverside, CA, USA
[2] Blue Marble Space Institute of Science, Seattle, WA, USA
[3] NASA Goddard Space Flight Center, 8800 Greenbelt Road, Greenbelt, MD 20771, USA
[4] Integrated Space Science and Technology Institute, Department of Physics, American University, Washington DC, USA
[5] Consortium on Habitability and Atmospheres of M-dwarf Planets (CHAMPs), Laurel, MD, USA
[6] Laboratoire de Météorologie Dynamique/IPSL, CNRS, Sorbonne Université, Ecole Normale Supérieure, PSL Research University, Ecole Polytechnique, 75005 Paris, France
[7] Laboratoire d'astrophysique de Bordeaux, Univ. Bordeaux, CNRS, B18N, allée Geoffroy Saint-Hilaire, 33615 Pessac, France

## Abstract

The habitable zone (HZ) is a fundamental concept used to guide our search for temperate exoplanets with stable surface liquid water that would sustain remotely detectable biosignatures. So far, most advanced 3D climate studies and prospective spectral simulations of HZ planets have focused on inner HZ planets that receive stellar radiation similar to Earth's and would consequently maintain relatively low (e.g., hundreds of parts per million) atmospheric $CO_2$. Here, we present a systematic investigation of outer HZ (OHZ) planets, which require high $CO_2$ for habitability, using the 3D Generic Planetary Climate Model (PCM). Our 3D climate results are compared with previous 1D predictions that show strong agreement between models. Simulated spectra of the modeled planets indicate that high overlying $CO_2$ levels affect remote signatures of habitability. We also determine substantial spectral differences that would distinguish habitable versus nonhabitable high-$CO_2$ worlds with future direct-imaging observations for solar analogs and with transmission and emission spectroscopy for M dwarfs. Our results demonstrate that 3D models that include $CO_2$ condensation are necessary to understand the potential planetary habitability and spectral impacts of clouds on terrestrial worlds in the OHZ.

## 1. Introduction

The search for life outside of the solar system is guided in part by the habitable zone (HZ), a conceptual region around a star where planets with the right atmospheric composition could maintain surface liquid water in contact with that atmosphere, potentially allowing for the accumulation of detectable spectroscopic biosignatures (J. F. Kasting et al. 1993; R. K. Kopparapu et al. 2013a; E. W. Schwieterman et al. 2018). The most accepted HZ paradigm assumes a $N_2$-$CO_2$-$H_2O$ atmosphere with the operation of a negative geochemical feedback called the carbonate-silicate cycle, where $CO_2$ is regulated over geologic time by the interplay between $CO_2$ outgassing and chemical weathering of $CO_2$ into carbonates, which are eventually subducted into the planetary interior (J. C. G. Walker et al. 1981; R. K. Kopparapu et al. 2013a). The greenhouse warming originates from a combination of $CO_2$ and $H_2O$, the latter of which is a condensable gas and on long timescales must be controlled by the $CO_2$ abundance. The carbonate-silicate cycle predicts that $CO_2$ concentrations will rise with increasing distance from the host star and decreasing instellation until very high-$CO_2$ levels are reached at the maximum greenhouse limit, which defines the outer habitable zone (OHZ) boundary and occurs when additional $CO_2$ becomes a net cooling rather than a net warming gas (J. F. Kasting et al. 1993; R. K. Kopparapu et al. 2013a). This feedback significantly extends the OHZ from 1.01 au (assuming Earth-identical atmospheric composition) to 1.67 au, where outer planets would need to maintain dense $CO_2$ atmospheres to support habitable surface conditions (J. F. Kasting et al. 1993). The levels of $CO_2$ required to warm a planet at the outer edge of the HZ range from several to tens of bars depending on the spectral type of the planet's host star (R. K. Kopparapu et al. 2013a). This is well beyond the $CO_2$ concentration that Earth's atmosphere has ever maintained, except perhaps for its earliest history (K. Zahnle et al. 2007; D. A. Stolper et al. 2016). The extent to which negative geochemical carbon feedbacks are a common occurrence in terrestrial exoplanets has not yet been empirically demonstrated (O. R. Lehmer et al. 2020), which will require a survey of nearby HZs and their spectral signatures (J. H. Checlair et al. 2019).

The propensity of a planet to enter or exit an ice-covered "snowball" state is dependent on the spectral energy distribution (SED) of its host star, which will differ across exoplanetary systems. For example, the large fraction of near-IR radiation received by planets orbiting M dwarfs renders them less susceptible to snowball episodes than planets orbiting stars with higher visible and near-UV output (A. L. Shields et al. 2013). When considering a fixed level of $CO_2$, M dwarf aquaplanets remain free of global ice cover with 25% less stellar flux than F dwarf aquaplanets and 19% less stellar flux than G dwarf aquaplanets (A. L. Shields et al. 2013). Distant planets that orbit beyond the HZ may exit this "snowball" state as a response to the increase in their host star's insolation as a function of the SED (A. L. Shields et al.

[8] NASA FINESST Fellow.

2014). A planet orbiting an M dwarf will require smaller insolation to initiate deglaciation than planets orbiting Sun-like stars (A. L. Shields et al. 2014; J. Checlair et al. 2017). As atmospheric $CO_2$ increases, this required insolation grows smaller, as M dwarf snowball planets demonstrate a stronger radiative response and will warm more efficiently than G dwarf snowball planets for the same increase in $CO_2$ (A. L. Shields et al. 2014). This equilibrium climate response of a planet to changes in insolation and stellar type has not yet been tested with a 3D climate model that includes $CO_2$ condensation physics.

Planets around M dwarfs are considered to have an advantage in some observation modes due to their favorable planet–star size ratios, high planet–star contrasts, relatively small angular separation, and enhanced signal-to-noise ratios. M dwarfs make up ∼70% of all stars and are expected to have an abundance of terrestrial-sized planets (G. D. Mulders et al. 2015) often found within the HZ (C. D. Dressing & D. Charbonneau 2015; S. Bryson et al. 2021). These worlds are expected to be tidally locked (S. H. Dole 1964; S. J. Peale 1977; J. F. Kasting et al. 1993; A. R. Dobrovolskis 2009), where one side of the planet always faces the star. Although their HZ boundaries have been calculated (e.g., R. K. Kopparapu et al. 2013a, 2014), the climates of synchronous rotators are not well approximated by 1D globally averaged models. 1D predictions for the OHZ have not yet been systematically examined in 3D as has been done for inner habitable zone (IHZ) planets, in part because of computation time. We expect that planets near the OHZ, having dense $CO_2$ atmospheres, should effectively transport heat to their nightsides and present a possible case for habitability. Evidence suggests that the inner planets, TRAPPIST-1 b and TRAPPIST-1 c, have thin or no atmospheres (T. P. Greene et al. 2023; S. Zieba et al. 2023). More broadly, recent James Webb Space Telescope (JWST) observations of highly irradiated rocky exoplanets indicate that many are consistent with bare-rock scenarios, although the presence of atmospheres cannot be definitively ruled out (L. Kreidberg & K. B. Stevenson 2025). For TRAPPIST-1 b specifically, JWST observations combined with 3D climate modeling suggest that the planet may either be a bare rock or sustain a $CO_2$-rich atmosphere with hazes (A. Maurel et al. 2025). However, coupled atmosphere-interior evolution models do not imply that these substantial atmospheres are unlikely for the outer planets (J. Krissansen-Totton 2023). As TRAPPIST-1 b and c are well inside the runaway greenhouse, prolonged exposure to high radiation likely led to atmospheric erosion, while the outer planets experienced shorter periods of extreme atmospheric loss and are much more likely to retain their atmospheres. The best opportunity to confirm these atmospheres remains with JWST transit observations of TRAPPIST-1 e and f (J. Krissansen-Totton 2023).

Potentially habitable planets around M dwarfs can be observed in transit with JWST to provide insight on whether or not these worlds have atmospheres, but the amount of observation time required to detect even the strongest spectral features for an Earth-like world is considerably high (C. V. Morley et al. 2017; V. S. Meadows et al. 2023; TRAPPIST-1 JWST Community Initiative et al. 2024). This is partially due to the effects of clouds dominating the spectrum (T. J. Fauchez et al. 2020; T. D. Komacek et al. 2020; G. Suissa et al. 2020), making $CO_2$ likely the most detectable gas for an Earth-like world (J. Lustig-Yaeger et al. 2019; T. J. Fauchez et al. 2020; D. Pidhorodetska et al. 2020). Additionally, stellar contamination may overshadow the planetary signal of terrestrial planets around M dwarfs (O. Lim et al. 2023; M. Radica et al. 2025). However, planets orbiting "Sun-like" (FGK) stars will be directly imaged with NASA'S Habitable Worlds Observatory (HWO). While the instrument capabilities are yet to be finalized, we expect the design to be motivated by previous mission concept reports including those of LUVOIR (The LUVOIR Team 2019) and HabEx (B. S. Gaudi et al. 2021). Other future facilities such as the Extremely Large Telescope (ELT) will also utilize direct imaging to study terrestrial exoplanets. In particular, the Planetary Camera and Spectrograph (PCS) is being developed for observations of temperate rocky planets around M and K dwarfs (M. Kasper et al. 2021).

The expansion of traditional HZ boundaries as a function of factors such as planetary eccentricity, obliquity, rotation rate, and surface albedo is also dependent on $pCO_2$ (e.g., E. Bolmont et al. 2016; S. R. Kane & S. M. Torres 2017; T. Jansen et al. 2019; I. Z. Palubski et al. 2020; P. Vervoort et al. 2022). Although other gases such as $CH_4$ and $H_2$ have been suggested to expand the HZ (e.g., S. Seager 2013; R. M. Ramirez 2019; M. Mol Lous et al. 2022), they lack fully explicated mechanisms for the strong abiotic negative feedback cycle to regulate greenhouse warming as a function of surface temperature and insolation. Hence, while consideration of alternative greenhouse gases is important, fully understanding traditional $CO_2$ greenhouse limits is a necessary first step and underpinned by empirical evidence for the carbonate-silicate cycle on Earth. Thus, that is the focus for this work. Determining how these variables affect each other is crucial to informing our ability to test HZ boundaries derived with simple 1D models with more advanced techniques. Previous attempts to characterize the HZ with global climate models (GCMs) have focused on the IHZ, where the amount of $CO_2$ is expected to be low due to high insolations and the negative feedback of the carbonate-silicate cycle (R. K. Kopparapu et al. 2016; Y. Fujii et al. 2017; G. Chaverot et al. 2023; M. Turbet et al. 2023). While some GCM studies looked at the climates of planets near the OHZ (R. D. Wordsworth et al. 2011; A. Paradise & K. Menou 2017; M. Turbet et al. 2017b, 2018; A. Paradise et al. 2021; K. Taniguchi et al. 2026), none have yet attempted to quantify the position of the OHZ as a function of stellar type, based on the maximum greenhouse limit.

The impact of clouds on the climates of terrestrial planets is very important as they can invoke the greenhouse effect by trapping the infrared radiation in the lower atmosphere or create an albedo effect by scattering incident stellar radiation back to space. The extent of the HZ around different types of stars is therefore dependent on the presence of clouds (M. S. Marley et al. 2013; J. Yang et al. 2014; R. K. Kopparapu et al. 2016, 2017; M. Turbet et al. 2021, 2023), especially for the OHZ boundary, which is influenced by the formation of $CO_2$ ice clouds. These $CO_2$ ice particles enable the scatter of thermal radiation back to the planets' surface, creating a scattering greenhouse effect (F. Forget & R. T. Pierrehumbert 1997). This effect could then increase the surface temperature above the freezing point of water (R. D. Wordsworth et al. 2011; F. Forget & S. Lebonnois 2013; M. Turbet et al. 2017b). Although the climatic effect of $CO_2$ ice clouds in $CO_2$-dominated atmospheres of terrestrial planets around different types of main-sequence dwarf stars has previously been explored (D. Kitzmann 2017), such a systematic study has not yet been performed with a 3D climate model, which is necessary to determine the locations of $CO_2$ clouds in such an atmosphere, particularly for synchronous rotators (R. D. Wordsworth et al. 2011; J. Leconte et al. 2013b).

The objective of this study is to understand how the results of a 3D climate model may (or may not) deviate from the previous 1D predictions while determining the spectral signatures of habitable OHZ worlds and how they differ from habitable worlds near the IHZ. Additionally, we establish how habitable worlds near the OHZ differ from nonhabitable high-$CO_2$ planets. We highlight a strong similarity between the behavior of water clouds that can form on the day or nightside (J. Yang et al. 2013; R. K. Kopparapu et al. 2016, 2017; M. Turbet et al. 2021, 2023) with those of $CO_2$ ice clouds (this work). Section 2 details the climate, spectral, and instrumental modeling methods considered in all simulations. Summarized results for average global temperatures can be found in Section 3, while an expanded discussion of the impacts of clouds, planetary parameters, and comparisons to previous 1D and 3D predictions is included in Section 4. Conclusions follow in Section 5.

## 2. Methods

### 2.1. 3D Climate Modeling

To test the limits of the OHZ as a function of increasing atmospheric $CO_2$, we ran a series of experiments using the Generic Planetary Climate Model (PCM), formerly known as the Generic LMD GCM, the only 3D GCM that includes $CO_2$ condensation physics. The Generic PCM has previously been developed and used to simulate a wide range of exoplanetary atmospheres ranging from that of temperate rocky planets (R. D. Wordsworth et al. 2011; J. Leconte et al. 2013a; M. Turbet et al. 2018; T. J. Fauchez et al. 2019) to warm mini-Neptunes (B. Charnay et al. 2015a, 2015b; M. Turbet et al. 2016; B. Charnay et al. 2021; C. Cadieux et al. 2024) and even hot Jupiters (L. Teinturier et al. 2024). Specifically, the model has been adapted to simulate water-rich planetary atmospheres with numerous applications including studying the runaway greenhouse on Earth (J. Leconte et al. 2013a; G. Chaverot et al. 2023) and water condensation on early Earth, Venus (M. Turbet et al. 2021), and exoplanets (M. Turbet et al. 2023). The Generic PCM has used the early Martian climate as a laboratory to run 3D simulations of thick, multibar, $CO_2$-dominated atmospheres (F. Forget & S. Lebonnois 2013; R. Wordsworth et al. 2013; R. D. Wordsworth et al. 2015; M. Turbet et al. 2017a, 2020). However, early Mars studies are calibrated specifically for Mars, which did not have a global ocean, and are not necessarily applicable to predicting the climates of larger, water-rich Earth-like terrestrial worlds in the OHZ. We build on these prior studies by exploring a larger $p\mathrm{CO_2}$-insolation parameter space to understand how these factors jointly modulate planetary climate for Earth-like worlds.

The Generic PCM includes complete radiative transfer that takes into account absorption and scattering by the atmosphere, the clouds, and the surface. The radiative-transfer calculations, which are based on the correlated-k approach (Q. Fu & K. N. Liou 1992), use 32 spectral bands in the thermal infrared and 35 in the visible domain. The opacity tables were calculated as described in M. Turbet et al. (2017a) using the HITRAN 2012 database (L. Rothman et al. 2013). The opacity tables also account for continuum absorptions using the HITRAN collision-induced absorption (CIA) database (T. Karman & J. M. Hutson 2019) and the MT_CKD v3.5 database (E. J. Mlawer et al. 2012). Subgrid-scale dynamical processes (turbulent mixing and convection) were parameterized as described in J. Leconte et al. (2013b). Moist convection was taken into account following a moist convective adjustment scheme that is originally derived from S. Manabe & R. T. Wetherald (1967) and was later generalized for water-dominated atmospheres in J. Leconte et al. (2013b). Relative humidity evolves freely and is limited to 100% (no supersaturation). In practice, when an atmospheric grid cell reaches 100% saturation and the corresponding atmospheric column has an unstable vertical temperature profile, the moist convective adjustment scheme is performed to obtain a stable, moist adiabatic lapse rate. When condensing, water vapor forms liquid water droplets and/or water ice particles, depending on the atmospheric temperature and pressure, forming clouds.

We used a fixed number of activated cloud condensation nuclei (CCNs) per unit of mass of air to determine —based on the amount of condensed material—the local, effective radii of $H_2O$ and $CO_2$ cloud particles (R. D. Wordsworth et al. 2011; M. Turbet et al. 2021). The effective radius is then used to compute the radiative properties and the sedimentation velocity of cloud particles. Water precipitation is divided into rainfall and snowfall, depending on the nature of the cloud particles, determined solely by the atmospheric temperature. Rainfall is parameterized to account for the conversion of cloud liquid droplets to raindrops by coalescence with other droplets (O. Boucher et al. 1995). Rainfall is considered to be instantaneous, that is it goes directly to the surface, but can evaporate while falling through subsaturated layers (D. Gregory 1995). Ice particles of both species were sedimented according to Stokes's law (F. Forget et al. 1999). The snowfall rate is calculated using the sedimentation velocity of particles, assumed to be equal to the terminal velocity that we approximate by a Stokes's law modified with a "slip-flow" correction factor (W. B. Rossow 1978).

The simulations presented in this paper were performed at a spatial resolution of $64 \times 48$ in longitude $\times$ latitude. The exact number of atmospheric layers and the model top pressure vary slightly depending on the surface pressure considered. The entire pressure field is fully dynamic including the surface boundary. The dynamical time step of the simulations is $\sim$120 s, and the radiative time step is 30 minutes, but there can be minor changes between simulations. The GCM includes condensation, evaporation, sublimation, and precipitation of water.

This study considers modeled planets around our Sun and TRAPPIST-1 to determine the influence of varying SEDs on atmospheric $CO_2$ in conjunction with other planetary parameters. Insolation values were varied to represent the boundaries of the OHZ (0.5–0.1 Seff) and $p\mathrm{CO_2}$ from 5 to 20 bar, consistent with estimates of early Mars and OHZ $CO_2$ estimates (R. K. Kopparapu et al. 2013a, 2013b, 2014). The Generic PCM has been used to study planets similar to TRAPPIST-1 f with $p\mathrm{CO_2}$ values up to 30 bar (R. D. Wordsworth et al. 2011; M. Turbet et al. 2018; T. J. Fauchez et al. 2019). This work intends to go beyond previous exploratory studies by performing a systematic analysis of the limits of the OHZ. As we expect planets to form hot due to their initial accretion energy and may also have early impact-induced greenhouses (K. J. Zahnle et al. 2020), simulations begin with a "warm start" where the initial surface temperature is equal to 300 K. Each modeled planet begins with Seff = 0.75; if the planet is frozen at such an insolation, further analysis was not deemed necessary at lower insolation values, but if the planet freezes between values considered in the parameter grid (Table 1), additional runs were deemed necessary to determine the minimum insolation required to keep the planet temperate. The planets are modeled with anoxic atmospheres containing only $CO_2$ and $H_2O$, which has a variable mixing ratio. Generic PCM can account for changes in ocean salinity by adjusting the freezing point of water (M. Turbet et al. 2016). However, when testing the range of ocean salinity from 5 to 50 ppt to consider estimates of salinity variations in Earth's history (L. P. Knauth 1998), the impact on the freezing point of water is negligible (reducing 1°–2° between 5 and 50 ppt) compared to the warming (or cooling) effects of adjusting $pCO_2$ by 1+ bars. Therefore, we set all salinity values to 35 ppt. Each simulation is an aquaplanet with a uniform slab ocean.

**Table 1**
Modeled Planetary Parameters

| $pCO_2$ (bar) | Seff | Obliquity (deg) | Day Length (hr) |
|---|---|---|---|
| Sun | | | |
| 5 | 0.25–0.5 | 23.5, 0 | 24 |
| 7.5 | 0.34–0.4 | 23.5, 0 | 24 |
| 10 | 0.25–0.39 | 23.5, 0 | 24 |
| TRAPPIST-1 | | | |
| 5 | 0.25–0.38 | 0 | Synchronous |
| 10 | 0.1–0.25 | 0 | Synchronous |
| 20 | 0.1–0.25 | 0 | Synchronous |

Assuming the Sun as the host star, planetary rotation periods are set to 24 hr, while obliquity is set to Earth's 23°5. All other planetary parameters are fixed at their present-day Earth values. When testing the influence of planetary obliquity on the average global temperature, the differences between 23°5 and 0° were found to be insignificant, varying by only 1–2 K. For this reason, we present only the 23°5 obliquity cases in this work.

Planets around TRAPPIST-1 are parameterized by TRAPPIST-1 f, a known planet in the OHZ that is tidally locked with 0° obliquity. While insolation and $pCO_2$ are varied, all other input parameters for TRAPPIST-1 f reflect those reported in E. Agol et al. (2021), where $M_\oplus = 1.039 \pm 0.031$, $R_\oplus = 1.045^{+0.013}_{-0.012}$, and $\rho$ (g cm$^{-3}$) = $5.02^{+0.14}_{-0.16}$. A full list of the parameter space explored can be found in Table 1.

### 2.2. Spectral Modeling

Transmission, emission, and reflected light spectra are simulated using the Planetary Spectrum Generator (PSG; G. L. Villanueva et al. 2018, 2022), an online radiative-transfer suite that computes synthetic spectra for a given planet. For climatological data taken from GCMs, PSG has a Global Emission Spectra (GlobES; T. J. Fauchez et al. 2025) application that aggregates and maps 3D temporal atmospheric data into a 2D projected observational grid. This grid is then provided to PSG's radiative-transfer modules, which are then used for all simulations. PSG considers the same stellar and planetary parameters as used by the GCM and has multiple scatterings fully included. The resolving power is defined as $R = \lambda/\Delta\lambda$, where $\Delta\lambda$ is the spectral bin width at the wavelength $\lambda$. Reflected light spectra for the Sun host are generated over a wavelength range of 0.2–2.0 $\mu$m at $R = 200$. Transmission spectra for TRAPPIST-1 planets are generated over a wavelength range of 0.6–12 $\mu$m at $R$= 200. Emission spectra are generated for both stellar types over a wavelength range of 4–20 $\mu$m at $R = 200$.

**Table 2**
Average Surface Temperatures for Modeled Planets around the Sun at 23.5 Obliquity

| $pCO_2$ (bar) | $S$ [$S_\oplus$] | Avg. Temperature (K) |
|---|---|---|
| 5 | 0.38 | < 250 |
| | 0.39 | 284 |
| | 0.40 | 294 |
| 7.5 | 0.37 | < 250 |
| | 0.38 | 294 |
| | 0.39 | 301 |
| 10 | 0.37 | < 250 |
| | 0.38 | 298 |
| | 0.39 | 308 |

## 3. Results

For modeled planets around the Sun, our model indicates that the OHZ boundary for a $pCO_2 = 5$ bar atmosphere as a function of insolation is at Seff = 0.39, while both the 7.5 and 10 bar atmospheres are Seff = 0.38 (Table 2). Any insolation lower than these did not reach an equilibrium state in the model, indicating a frozen planet. For these nonhabitable planets, surface temperatures drop below the condensation temperature of $CO_2$ (at least on the nightside), causing the atmosphere to collapse, bringing the surface temperature below the melting point of water. In such cases, the pressure is steadily dropping, and is presumed to reach 0 bar if continued to run indefinitely. To conserve computing time, we end these simulations before this point is reached, when the planet is completely glaciated with 0% liquid $H_2O$, and the average surface temperature drops below 250 K.

For simulations with TRAPPIST-1, tidally locked planets are better able to maintain fractional habitability (i.e., partially glaciated states) at low insolations near the OHZ boundary (Table 3). For $pCO_2 = 5$ bar atmospheres, these include Seff = 0.28–0.35 (Figure 1). When using a 3D GCM with an active carbon cycle, tidally locked planets in the HZ do not experience limit cycling between warm and snowball states (J. H. Checlair et al. 2019). Instead, they settle into an "eyeball" state, which is primarily frozen with the exception of an unglaciated substellar region (R. T. Pierrehumbert 2011). We show here that, as $pCO_2$ increases, the OHZ boundary is pushed out, allowing planets with insolations as low as Seff = 0.19 where $pCO_2 = 10$ bar to be potentially habitable. This indicates that the HZs of M dwarf systems may be underestimated by 1D models when high-$CO_2$ values (>10 bar) are considered.

**Table 3**
Average Surface Temperatures and Their Open Ocean Fraction for Modeled Planets around TRAPPIST-1

| $pCO_2$ (bar) | $S$ [$S_\oplus$] | Avg. Temperature (K) | OOF (%) |
|---|---|---|---|
| 5 | 0.27 | < 250 | 0 |
| | 0.28 | 259 | 16 |
| | 0.29 | 263 | 23 |
| | 0.30 | 265 | 27 |
| | 0.31 | 266 | 30 |
| | 0.32 | 270 | 40 |
| | 0.33 | 275 | 60 |
| | 0.34 | 279 | 94 |
| | 0.35 | 285 | 100 |
| | 0.38 | 301 | 100 |
| 10 | 0.18 | < 250 | 0 |
| | 0.19 | 271 | 71 |
| | 0.20 | 282 | 100 |
| 20 | 0.15 | < 250 | 0 |
| | 0.20 | 337 | 100 |

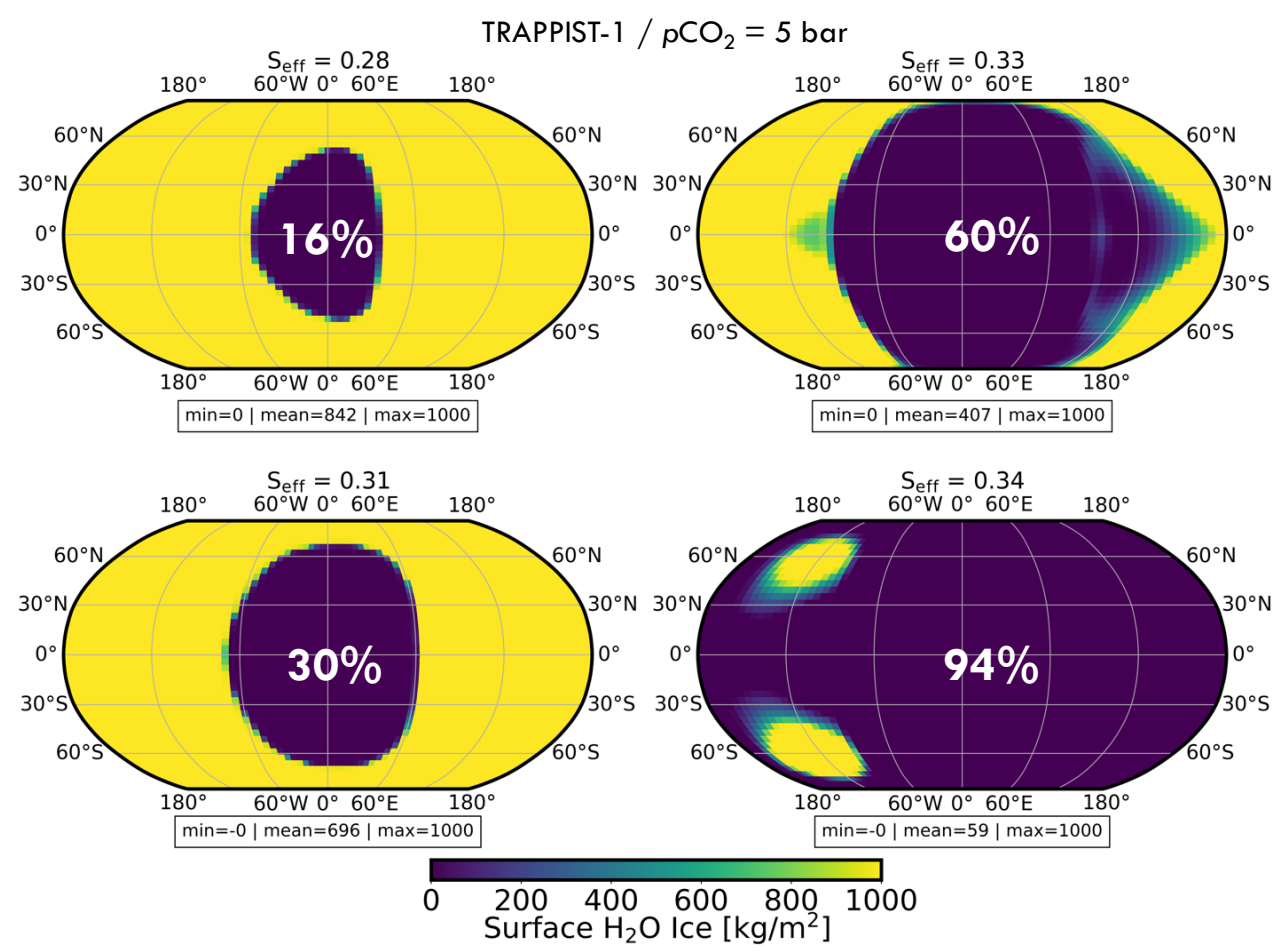

**Figure 1.** Surface water ice coverage for planets modeled around TRAPPIST-1 whose insolation values range from Seff = 0.28–0.32 where $pCO_2 = 5$ bar. The open ocean fraction (OOF) is shown as a percentage (white) for each simulation. Corresponding surface temperatures are noted in Table 3.

### 3.1. Clouds

When considering cloud distribution for cold nonhabitable planets around the Sun and TRAPPIST-1, $CO_2$ clouds strongly dominate (due to widespread $CO_2$ condensation) and overshadow the influence of $H_2O$ clouds (Figure 2). GCMs that do not include $CO_2$ condensation will not include this effect, creating considerable differences in cloud coverage. Figure 3 illustrates this result by showing combined $H_2O + CO_2$ cloud coverage and the resulting surface temperatures. By combining these effects and looking at the outgoing longwave radiation, we also find that the $CO_2$ clouds block thermal radiation and prevent it from escaping, while the lack of $CO_2$ clouds for habitable worlds has the opposite effect (Figure 4). Furthermore, the absence of clouds reduces the planetary albedo, making warmer planets much darker than frozen worlds, where surface $CO_2$ ice is highly reflective. This is consistent with previous work suggesting that water clouds have a greater influence on the planetary albedo for planets near the IHZ compared to the OHZ (F. Selsis et al. 2007). Increased surface temperatures near the IHZ lead to stronger convection and higher evaporation rates that result in more extensive water cloud coverage. These clouds typically form at higher altitudes, where they are effective at reflecting incoming stellar radiation and increasing the planetary albedo. In contrast, lower instellations near the OHZ can result in cooler surface temperatures, which would reduce evaporation and convection rates. In turn, this may result in less extensive water cloud coverage, as shown by our simulations. Stratiform clouds that form in these colder regions tend to be lower and thinner than those near the IHZ, which may reduce their effectiveness in reflecting stellar radiation.

Our simulations of cold nonhabitable planets orbiting TRAPPIST-1 reveal that $CO_2$ ice clouds are concentrated on the nightside (Figures 2 and 3). These clouds trap outgoing radiation and create a greenhouse effect that raises the nightside temperature above what it would be in their absence (although not enough to approach habitable conditions). On the dayside, strong heating drives stratospheric winds, transporting warm air to the nightside, similar to the behavior of water vapor described by Y. Fujii et al. (2017). When considering only water vapor on hot planets, M. Turbet et al. (2023) found that these warm air parcels cool as they travel to the nightside, losing heat through longwave emission. This cooling lowers the temperature to the saturation point of water vapor, leading to large-scale condensation and the formation of stratospheric clouds located on the nightside and at the poles.

The $CO_2$ cloud asymmetry between the day and nightside of these worlds and their net warming effect indicate a strong correlation to the $H_2O$ cloud asymmetry found in M. Turbet et al. (2021) when modeling early Venus. In M. Turbet et al. (2021), the formation of clouds on the nightside is driven by condensation of water vapor as the dominant gas, producing a strong greenhouse effect that reduces thermal cooling of the space. On the planetary dayside, water vapor strongly absorbs shortwave radiation and warms the upper layers of the atmosphere, further inhibiting the convection. The resulting atmospheric warming breaks moist convection and inhibits dayside cloud formation. As insolation increases, nightside temperatures rise, leading to a reduction in cloud coverage and resulting greenhouse warming. While upper-level shortwave heating suppresses convection in temperate planets, intense surface cooling plays the same role for planets in a snowball state. $CO_2$ similarly contributes to the moist convection subroutine through the inclusion of $CO_2$ condensation physics. Our results indicate that this same pattern applies to $CO_2$ clouds in a $CO_2$-dominated atmosphere for nonhabitable planets around M dwarfs.

For planets that enter a snowball state, extensive ice coverage significantly increases planetary albedo (Figure 4), reflecting incoming stellar radiation and lowering dayside surface temperatures. Cold surface temperatures prevent warm air from rising, further suppressing convection. Although M. Turbet et al. (2021) shows that this mechanism can be found in a wide range of planetary rotation periods, cloud and atmospheric circulation feedbacks may vary with rotation period (T. Jansen et al. 2019), highlighting the need for a sensitivity study to confirm this effect.

Beyond "eyeball" climates, other studies have shown that the deglaciated substellar region of temperate, tidally locked planets around M dwarfs displays a lobster-like pattern of surface $H_2O$. Previously, this pattern has only been seen in simulations that include ocean heat transport (OHT), which extends the area of open water along the equator (Y. Hu & J. Yang 2014; A. D. Del Genio et al. 2019). However, the Generic PCM does not include a dynamic ocean in the version of the model used here, but we note that the $CO_2$ clouds in cold simulations form a similar pattern (Figure 2). Yet, this pattern does not become apparent until the average surface temperature of the planet is below the condensation temperature of $CO_2$. The recent implementation of an improved dynamic slab ocean model (S. Bhatnagar et al. 2026) provides an opportunity to further investigate these observations.

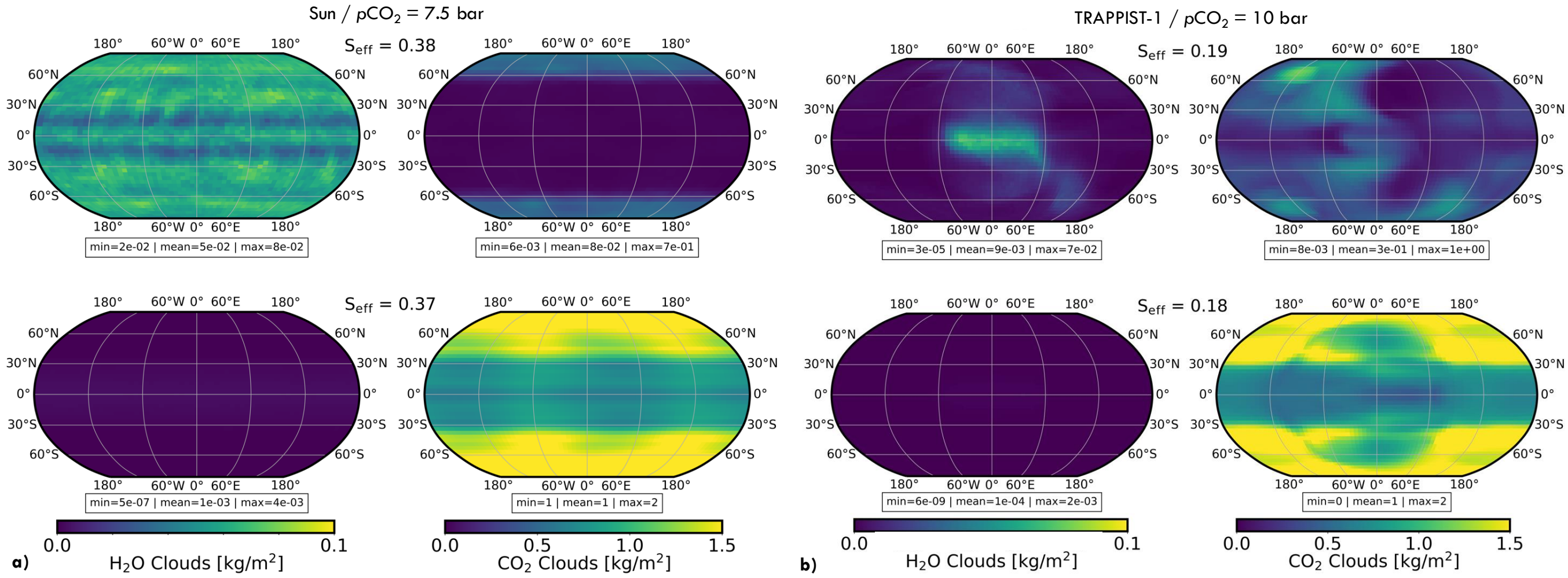

**Figure 2.** (a) $H_2O$ cloud (left column) and $CO_2$ cloud (right column) coverage for OHZ "edge cases" around the Sun where $pCO_2 = 7.5$ bar. The top row indicates the minimum insolation that converges to a potentially habitable planet in these conditions. The bottom row shows the slightly lower insolation where the planet continues to cool over time ultimately becoming frozen. (b) $H_2O$ cloud (left column) and $CO_2$ cloud (right column) coverage for OHZ edge cases around TRAPPIST-1 where $pCO_2 = 10$ bar. The top row indicates the minimum insolation value that results in a potentially habitable planet whose liquid $H_2O$ fraction = 100%. The bottom row shows a lower insolation value where the planet is frozen.

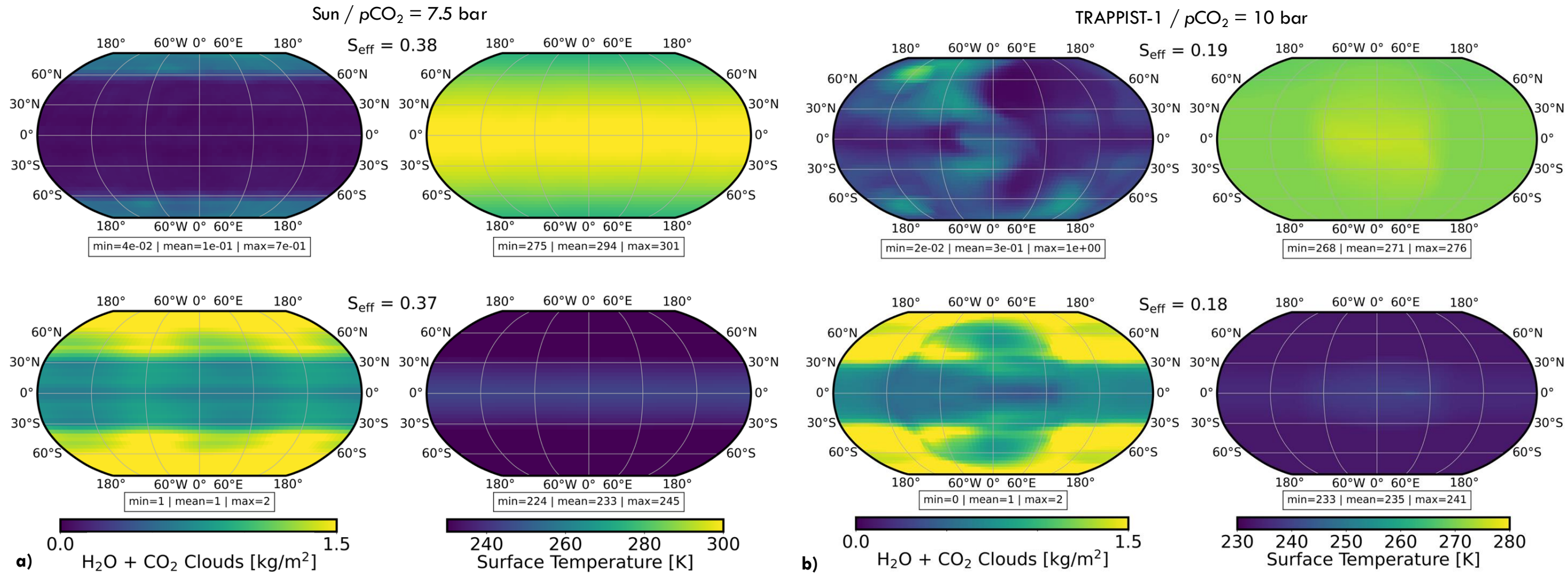

**Figure 3.** $H_2O$ cloud coverage combined with $CO_2$ cloud coverage (left column) and average surface temperatures (right column) for edge cases around the Sun (a) and TRAPPIST-1 (b) shown in Figure 2. In all simulations, $CO_2$ clouds strongly dominate the effects of $H_2O$ clouds.

Although OHT broadens the region of the surface liquid $H_2O$, our $CO_2$ clouds (or lack thereof) extend the area of the planet with an increase in surface temperature. Unfortunately, the temperature values where this pattern occurs (<250 K) are not high enough to form potential areas of habitability where the planet is otherwise frozen. We note that these temperatures are just below the freezing temperature of saturated NaCl brines, but brines with other types of salts such as $FeCl_3$ and $NaClO_4$ would be stable (V. F. Chevrier & R. A. Slank 2024), so in principle, we could not exclude the possibility of local habitable refugia.

Figure 5 shows a comparison between simulations performed with and without $CO_2$ condensation for a habitable OHZ planet around a Sun-like star. When $CO_2$ condensation is included, widespread $CO_2$ cloud formation limits greenhouse warming and maintains lower global temperatures. In simulations that neglect $CO_2$ condensation, this cooling mechanism is absent, causing average surface temperatures to be ∼10–20 K warmer. This result demonstrates that neglecting $CO_2$ condensation can lead to an overestimation of warming in high-$CO_2$ atmospheres.

### 3.2. Spectral Signatures of Planets in the OHZ

Figure 6 shows the reflected light and thermal emission spectra for the modeled OHZ boundary of the Sun (red) compared to a colder planet that does not reach equilibrium (blue). In reflected light, habitable and nonhabitable OHZ planets are distinguished by broadband albedo and the presence or absence of $H_2O$ absorption bands. Habitable planets are substantially darker than nonhabitable planets due to the darkness of liquid water oceans contrasted with the high reflectivity of $H_2O$ and $CO_2$ ices.

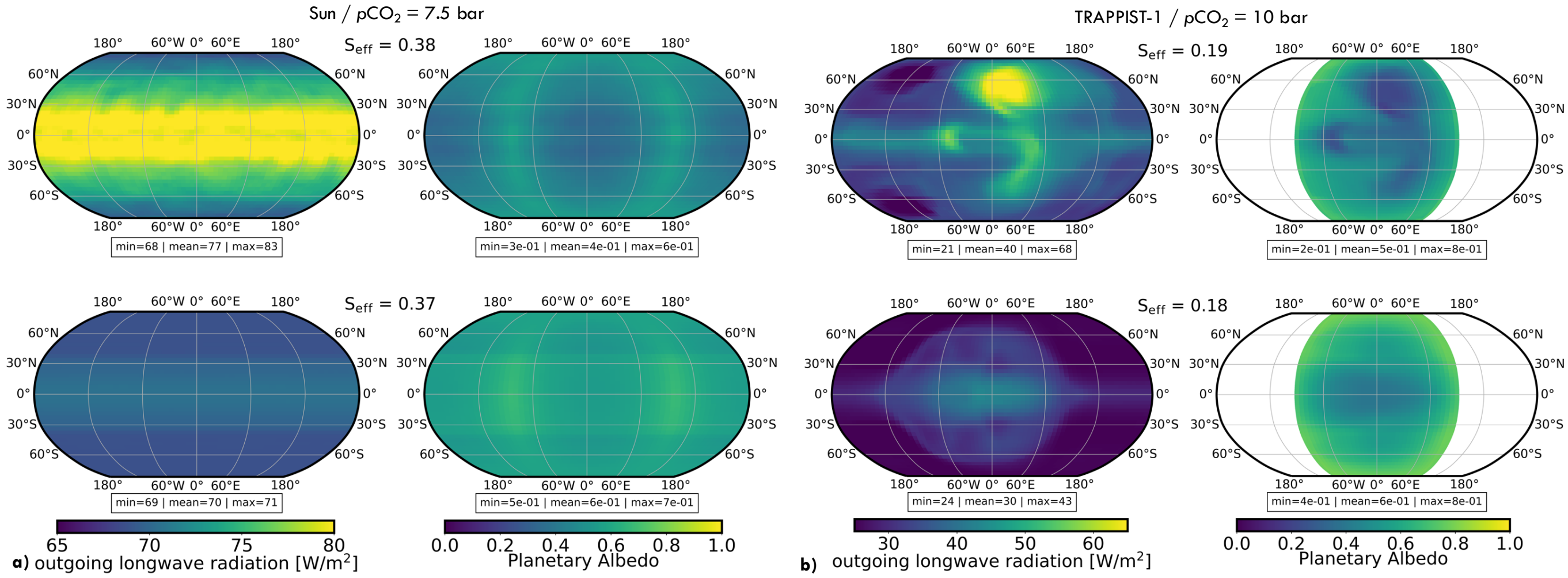

**Figure 4.** For edge cases around both stars, the outgoing longwave radiation (left column) for the cold planet scenarios is blocked by the presence of $CO_2$ clouds, which dominate the influence of $H_2O$ clouds (Figure 3). For the potentially habitable scenarios, the lack of $CO_2$ clouds has the opposite effect. Habitable planets are substantially darker than nonhabitable planets (right column) due to the lower albedo of the ocean compared to the highly reflective $H_2O$ and $CO_2$ ices and differences in cloud cover.

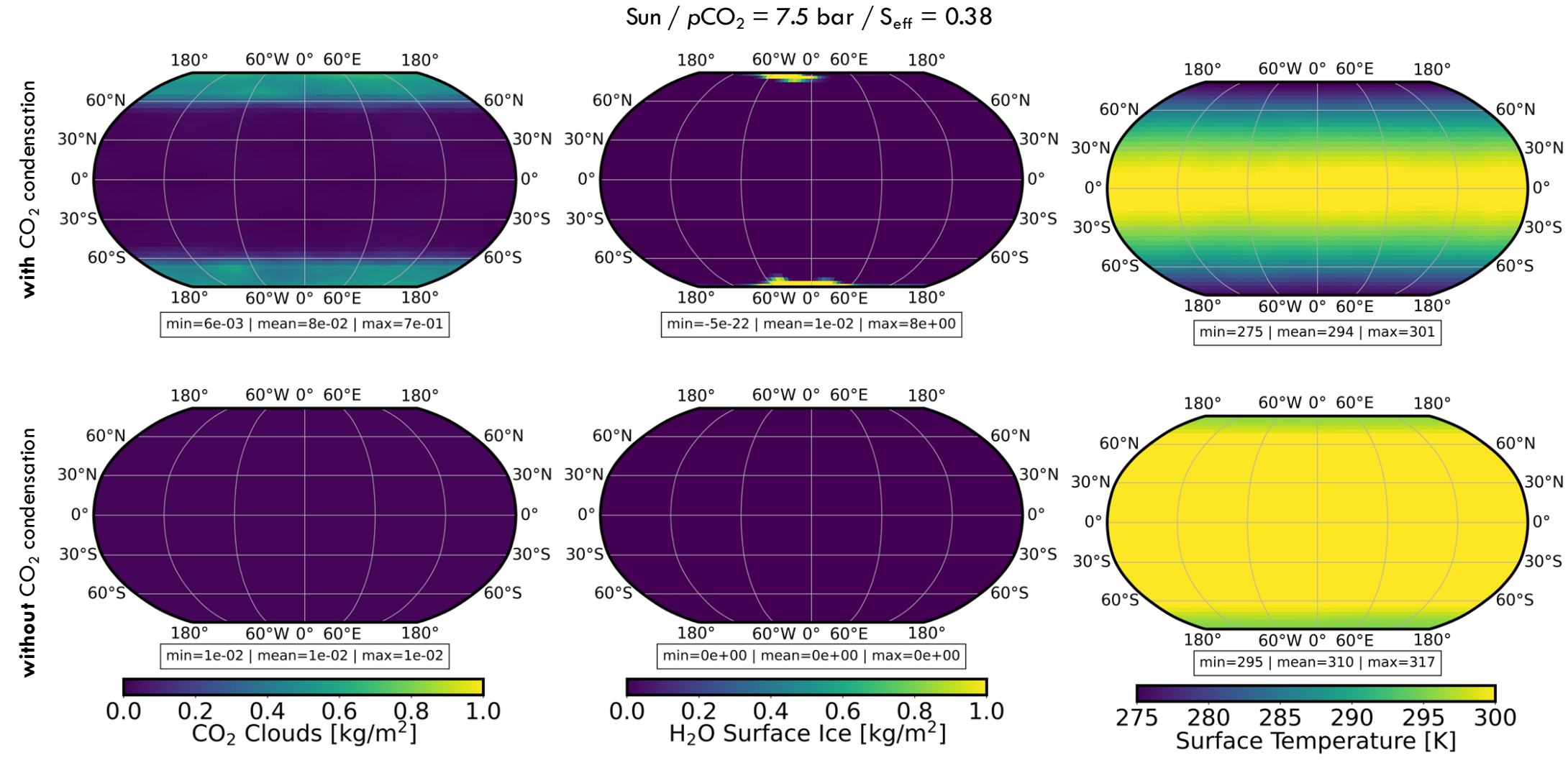

**Figure 5.** $CO_2$ cloud coverage (left), $H_2O$ surface ice (middle), and average surface temperatures (right) for a warm planet around the Sun where $pCO_2 = 7.5$ bar and Seff = 0.38. The top row shows the planet with $CO_2$ condensation included, as it is for all other simulations within this work, aside from the bottom row, where simulations do not include $CO_2$ condensation. In such cases, the average surface temperature may increase ∼10–20 K.

Frozen planets are dominated by $CO_2$ features from the overlying atmosphere and surface ice. The habitable planets contain $H_2O$ features including strong features at 0.94, 1.15, and 1.4 $\mu$m. However, these features are measured against a much darker continuum brightness. As a result, muted $H_2O$ lines in this region may be challenging to detect, requiring more integration time to characterize habitable planets than frozen ones.

While UV–vis–near-IR direct imaging would not be possible for a system as compact and distant as TRAPPIST-1 due to inner-working-angle considerations, transmission and emission spectra can be used to study modeled planets in the OHZ around TRAPPIST-1 (and extrapolated to similar systems). Figure 7 shows these spectra for a range of potentially habitable worlds while comparing them to a colder planet. Clouds are highly opaque in the visible and infrared, causing the spectral continuum to be elevated above the cloud deck where the atmosphere is semitransparent. A higher continuum reduces the transit depth of molecular absorption features, diminishing the detectability of $H_2O$ vapor, as it is mainly cold trapped below the cloud deck. For planets where the surface liquid water fraction varies between 0% and 100%, this is particularly important, as the detection of $H_2O$ vapor is critical in determining planetary habitability. In contrast, for planets whose liquid water fraction is 100% (such as Seff = 0.38 for $pCO_2 = 5$ bar), the atmosphere is warm enough to mitigate this issue, allowing for more reliable detection of water vapor. Here, the disappearance of $CO_2$ clouds and resulting changes in atmospheric temperature and scale height reduce the baseline continuum.

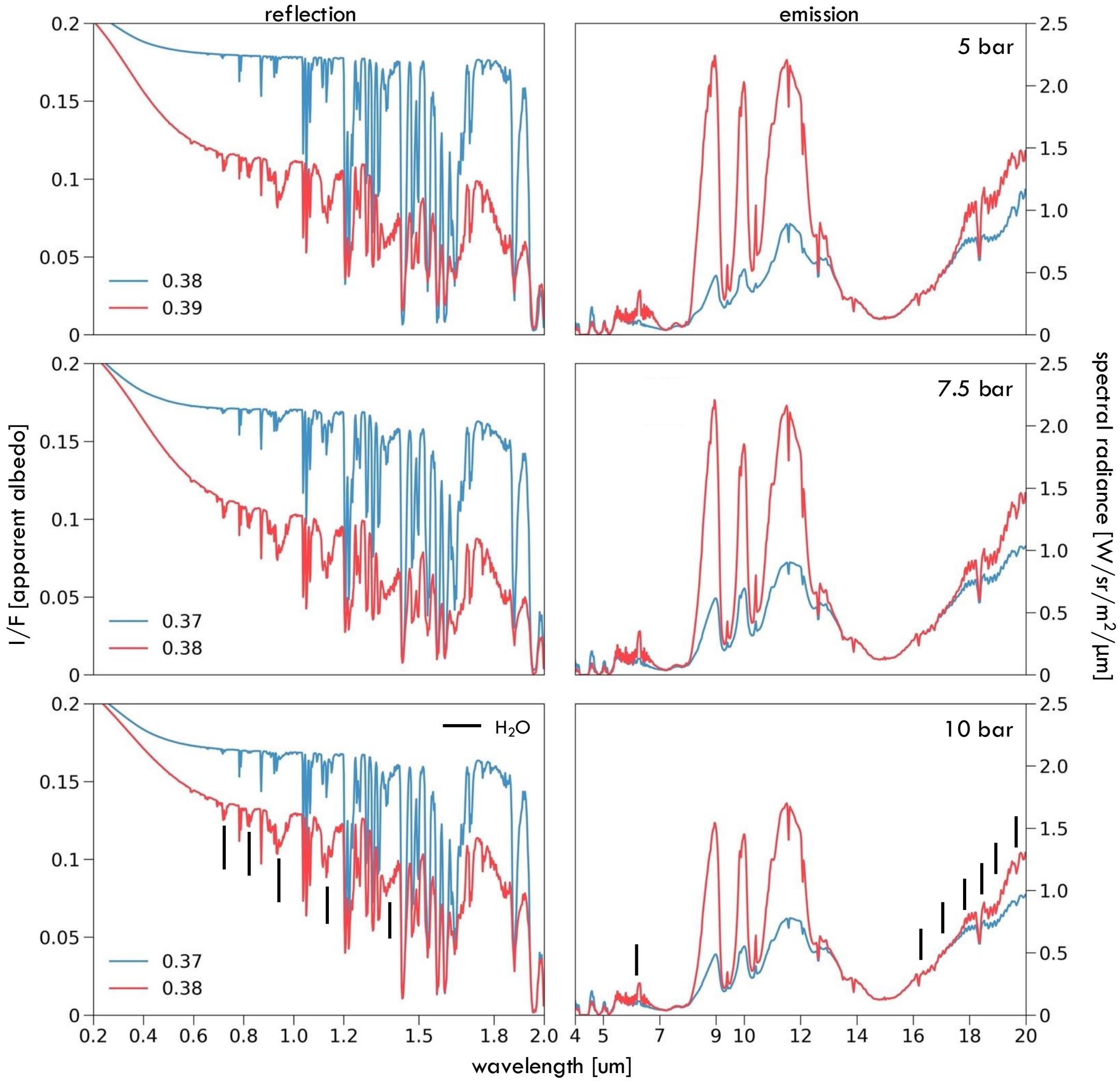

**Figure 6.** Modeled reflection (left column) and emission (right column) spectra for planetary "edge cases" around the Sun. Each row represents a different $pCO_2$ value of 5, 7.5, and 10 bar, respectively. Blue spectral lines indicate the highest insolation for each $pCO_2$ considered that does not reach equilibrium. Red spectral lines denote the lowest insolation that reaches equilibrium and is considered to be a habitable planet. $H_2O$ features are noted with black lines. All unlabeled features correspond to $CO_2$. Temperature values that correspond to these insolation values can be found in Table 2.

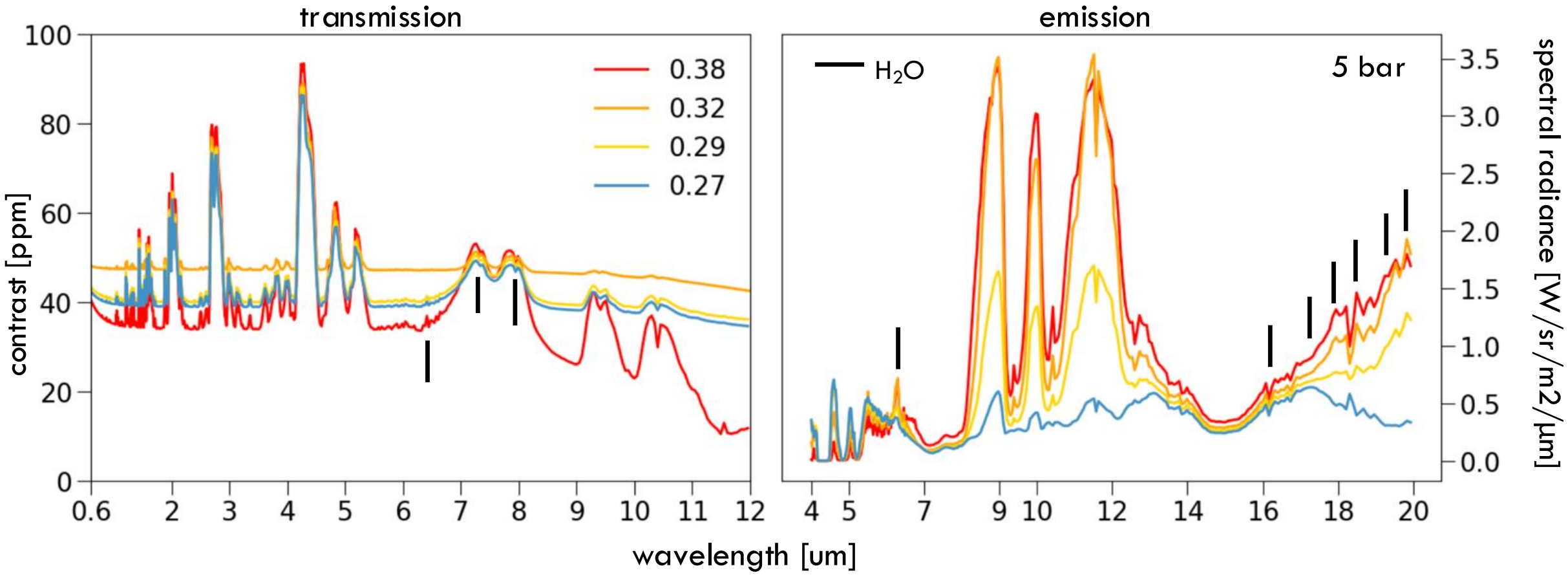

**Figure 7.** Modeled transmission (left column) and emission (right column) spectra for planetary "edge cases" around TRAPPIST-1 at $pCO_2 = 5$ bar. Blue spectral lines indicate the highest insolation considered that does not reach equilibrium. Red, orange, and yellow spectral lines denote insolation values that reach equilibrium and are considered to be habitable planets. Temperature values that correspond to these insolation values can be found in Table 3.

Planetary thermal emission spectra could be obtained by future space-based interferometers such as the LIFE mission concept (S. P. Quanz et al. 2018, 2022; E. Alei et al. 2022; D. Angerhausen et al. 2024). Direct imaging in the thermal infrared could also be feasible for select systems using 30 m class ground-based telescopes (Y. Fujii et al. 2018;

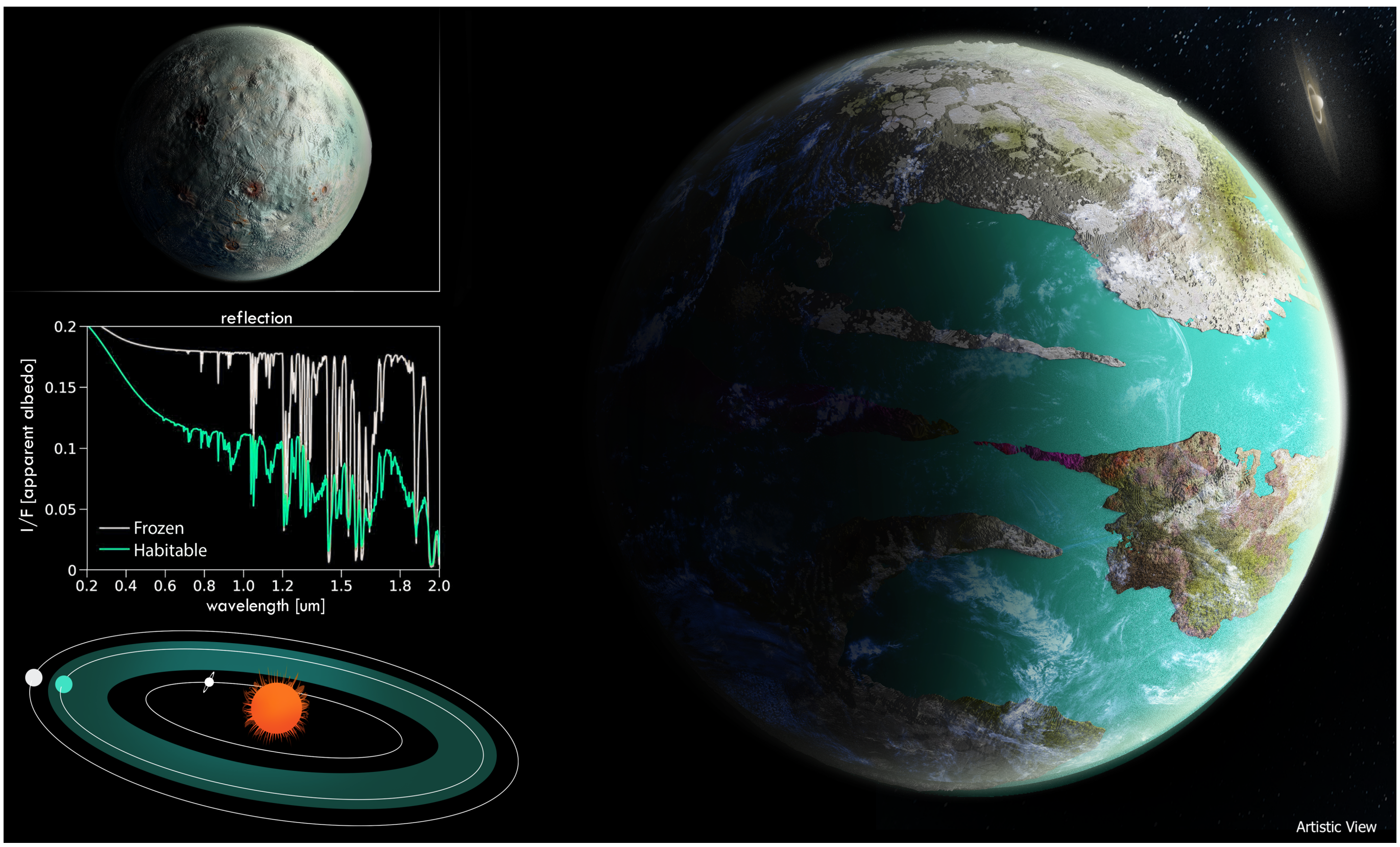

**Figure 8.** Simulated climates and observables for $CO_2$-rich planets near the outer edge of the habitable zone. The illustration on the right depicts a potentially habitable world orbiting a Sun-like star. The teal point in the lower left section of the panel marks the orbital location of such a planet within the OHZ, and the corresponding reflected light spectrum that could be observed via direct imaging is shown in teal in the middle, left section of the panel. The upper left section of the panel shows an illustration of the same planet in a globally frozen state located just beyond the habitable zone. Its orbital position (lower left) and reflected light spectrum (middle) are shown in white.

M. Lopez-Morales et al. 2019). As the specifications and parameters of future reflected light and thermal emission observations have not yet been defined, we do not attempt to quantify the detectability of specific atmospheric features. However, we believe that these simulations will serve as valuable input for future studies focused on planetary characterization of planets throughout the HZ, including the OHZ. We note that any future observations would have to contend with lower thermal emission signatures for habitable OHZ planets than for more Earth-like targets. To briefly quantify, the thermal emission of our habitable $pCO_2 = 5$ bar atmosphere (Figure 6) is about a factor of 3 dimmer than Earth at 11–12 $\mu$m (T. D. Robinson & C. T. Reinhard 2018).

## 4. Discussion

To summarize the climatic and observational implications of high-$CO_2$ planets near the OHZ, we present a conceptual illustration in Figure 8. The figure contrasts a potentially habitable world (right) in the OHZ with a globally frozen planet (upper left) located just beyond the OHZ. Frozen planets exhibit higher apparent albedos driven by reflective $CO_2$ ice, while habitable planets with surface oceans appear darker and exhibit spectral signatures of atmospheric water vapor. These differences highlight how future direct-imaging observations could distinguish temperate OHZ planets from colder high-$CO_2$ worlds that lie beyond the HZ.

### *4.1. Comparison to 1D Predictions*

Our simulations of high $pCO_2$ atmospheres for planets orbiting the Sun demonstrate strong consistency with the 1D predictions of the OHZ (J. F. Kasting et al. 1993; R. K. Kopparapu et al. 2013b, 2014). The maximum greenhouse limit for each star corresponds to a minimum Seff for the $pCO_2$ that provides maximum warming (i.e., increased planetary albedo with $pCO_2$ does not net counteract greenhouse warming). The ideal $pCO_2$ corresponding to the maximum greenhouse limit will vary depending on the SED of the host star, with higher planetary albedos and therefore lower maximum $pCO_2$ values for earlier-type stars, and much shallower "turnover" for late-type stars (R. K. Kopparapu et al. 2013a, 2014). The OHZ maximum greenhouse limit for the Sun found by R. K. Kopparapu et al. (2013a, 2013b) is 1.67 au where Seff = 0.36 and where atmospheric $pCO_2 = 7.5$ bar, while our results suggest that it is closer to 1.64 au with Seff = 0.37 (Table 4). This is a remarkable consistency given the numerous differences in model assumptions and complexity, validating past HZ assumptions for Sun-like stars.

1D models predict the $pCO_2$ versus Seff curve for cooler stars is relatively flat at high $pCO_2$, suggesting habitable planets orbiting cool stars at OHZ limits cannot have "too much" $CO_2$ to remain habitable at the limiting Seff, unlike planets orbiting Sun-like stars. We tested this relationship between $pCO_2$ and Seff for TRAPPIST-1, an ultracool dwarf, with our 3D model. Our results indicate a habitability boundary of Seff = 0.28 for $pCO_2 = 5$ bar, compared to the 1D prediction of Seff = 0.24 from R. K. Kopparapu et al. (2013a, 2014). At $pCO_2$ = 10–20 bar, our results suggest a limit around Seff = 0.16–0.20, whereas R. K. Kopparapu et al. (2013b, 2014) predicts a slightly more conservative value of Seff = 0.22. This difference may be attributable to differences in cloud feedback, radiative transfer, or atmospheric dynamics captured in 3D models that are not considered in 1D calculations. While our results are generally consistent with previous 1D studies, the inclusion of multidimensional climate effects may extend the OHZ boundary for late-type stars, especially ultracool dwarfs.

**Table 4**
OHZ Limits from Our 3D GCM

| $pCO_2$ (bar) | Model | Max Greenhouse [Seff] | Max Greenhouse (au) |
|---|---|---|---|
| Sun | | | |
| 5 | This work | 0.39 | 1.60 |
| | R. K. Kopparapu et al. (2014) | 0.38 | 1.62 |
| 7.5 | This work | 0.37 | 1.64 |
| | R. K. Kopparapu et al. (2014) | 0.36 | 1.67 |
| 10 | This work | 0.37 | 1.64 |
| | R. K. Kopparapu et al. (2014) | 0.36 | 1.67 |
| TRAPPIST-1 | | | |
| 5 | This work | 0.28 | 0.0432 |
| | R. K. Kopparapu et al. (2013a, 2014) | 0.24 | 0.0466 |
| 10 | This work | 0.19 | 0.0524 |
| | R. K. Kopparapu et al. (2013a, 2014) | 0.22 | 0.0487 |
| 20 | This work | 0.16–0.20 | 0.0511–0.0571 |
| | R. K. Kopparapu et al. (2013a, 2014) | 0.22 | 0.00487 |

While equilibrium temperature has formerly been used as a metric to determine potential habitability, W. J. Borucki et al. (2011), N. M. Batalha et al. (2013), and J. Leconte et al. (2013b) demonstrate that 1D models cannot predict nightside temperatures or mean temperatures accurate enough to draw strong conclusions around these planetary climates. Modeled planets around TRAPPIST-1 show many instances where habitability can be overlooked by 1D models. An analysis of the open ocean fraction (OOF) indicates strong examples of fractional habitability. In many cases, tidally locked planets whose global average temperatures are lower than 273 K may still have up to 40% OOF (Table 3). As there is currently no lower threshold value agreed upon for a fractionally habitable planet, we include all simulated results ranging from 0% to 100%.

### *4.2. Comparison to 3D Simulations*

Previous attempts to understand the potential habitability of TRAPPIST-1 f with a 3D GCM showed that, at 0.382 Seff, TRAPPIST-1 f can maintain a global mean surface temperature of 284 K with $pCO_2$ = 2 bar and 334 K with $pCO_2$ = 5 bar (E. T. Wolf 2018). Although model instabilities prevented us from exploring scenarios where $pCO_2 < 5$ bar, we note agreement with the conclusions made for $pCO_2$ = 5 bar, where our planet is habitable, but our surface temperature is lower (301 K compared to 334 K). One distinction between our study and that of E. T. Wolf (2018) is the inclusion of $CO_2$ condensation physics in our 3D climate model. Simulations in E. T. Wolf (2018) do not account for $CO_2$ condensation, potentially overestimating greenhouse warming at high atmospheric $pCO_2$. Our results indicate that, when $CO_2$ condensation physics are included, atmospheric collapse occurs at lower average temperatures, and OHZ boundaries derived from models neglecting $CO_2$ condensation may overestimate planetary habitability. Additionally, the planetary parameters for TRAPPIST-1 f have since been updated to reflect those in E. Agol et al. (2021), with the most significant change being made to planetary mass, increasing from 0.68 to 1.039 $M_\odot$. Lastly, the opacity tables in E. T. Wolf (2018) are sourced from HITRAN 2004 (L. S. Rothman et al. 2005), while ours are taken from HITRAN 2012 (L. Rothman et al. 2013) along with additional databases for CIAs and other continuum absorptions. Additionally, K. Taniguchi et al. (2026) investigated atmospheric collapse in $CO_2$-dominated atmospheres on a tidally locked planet using Generic PCM with planetary parameters representative of TRAPPIST-1f. Consistent with our results, they found that $CO_2$ condensation and atmospheric collapse play a fundamental role in determining the climates of cold, $CO_2$-rich terrestrial planets and should therefore be explicitly included in 3D climate simulations. While atmospheric collapse is often considered detrimental to habitability, K. Taniguchi et al. (2026) demonstrates that local dayside liquid water can persist after collapse as the reduction in atmospheric mass weakens day–night heat transport. Our studies help provide a more comprehensive picture of the climates of cold $CO_2$-dominated terrestrial planets by connecting the physical processes that lead to atmospheric collapse with their implications for HZ boundaries and observable atmospheric signatures.

Each simulation in this study represents an aquaplanet with a uniform slab ocean and no OHT. While this approach simplifies the model, future simulations may benefit from the addition of a dynamic ocean component that has a dynamic sea-ice model (M. J. Way et al. 2018; J. Yang et al. 2019), since sea-ice drift can strongly reduce OOF (J. Yang et al. 2020), and OHT can significantly influence the thermal structure and potential habitability of various terrestrial aquaplanets (M. J. Way et al. 2017, 2018; A. D. Del Genio et al. 2019). Efficient OHT can redistribute heat more evenly, potentially extending the OHZ by preventing polar regions from freezing. However, in the case of our simulations, which are characterized by a $CO_2$-dominated, high-pressure atmosphere, the influence of OHT may be less critical. High atmospheric $pCO_2$ enhances the atmospheric heat capacity and the ability to transport heat efficiently even in the absence of OHT. In contrast, for planets with low-pressure, $N_2$-dominated atmospheres, the lack of OHT could present a significant limitation as atmospheric circulation would not be an effective mechanism for heat transport. Further studies incorporating dynamic ocean models would help refine these conclusions, particularly for scenarios involving lower-pressure or $N_2$-dominated atmospheres.

The representation of convection in GCMs plays an important role in shaping the climate of tidally locked exoplanets. While a moist convective adjustment scheme is computationally efficient and widely used in exoplanet climate modeling, it presents limitations compared to more advanced schemes such as mass-flux parameterizations and convection-resolving models (CRMs). A simplified convection adjustment scheme has been shown to lead to a $> 60\%$ decrease in cloud albedo, increasing the mean dayside temperature by $\sim$10 K for TRAPPIST-1 e and Proxima Centauri b compared to a mass-flux parameterization (D. E. Sergeev et al. 2020), which is implemented in GCMs including ExoCAM, ROCKE-3D, UM, and MITGCM. However, both convection regimes still indicate habitable conditions for each planet (D. E. Sergeev et al. 2020). When considering potentially Earth-like worlds, GCMs may overestimate planetary heat transport efficiency and cloud coverage (D. E. Sergeev et al. 2020; M. Lefèvre et al. 2021). While we do not expect our overall findings to change based on the selected convection scheme, we note that future studies should consider higher CRMs to refine our climate predictions, particularly for interpreting telescope observations (D. E. Sergeev et al. 2020; M. Lefèvre et al. 2021).

The Generic PCM employs a two-stream radiative-transfer scheme, which may result in an overestimation of $CO_2$ cloud warming, as discussed by D. Kitzmann (2016). Although more sophisticated radiative-transfer schemes may be better suited for accurately capturing the effects of $CO_2$ clouds, the Generic PCM is uniquely suited to this study given the inclusion of $CO_2$ condensation physics. Further investigation using a more advanced radiative-transfer scheme would be a valuable direction for future work.

### *4.3. Planetary Parameters*

The climate of planets is affected by the reflective properties of their surfaces. The results presented in this work are based on simulations of aquaplanets, where the entire surface is covered by water. Water has a low albedo and high heat capacity, making this a somewhat optimistic assumption. Planets dominated by land are generally cooler and more reflective than those with higher fractional ocean coverage, although the magnitude of this effect is dependent on host star spectral type (A. J. Rushby et al. 2019). Changes in land fraction can result in differences of up to 20 K in globally averaged surface temperature (E. Macdonald et al. 2022). Nonetheless, ocean-dominated worlds are one plausible outcome of planet evolution, as Earth itself was likely free of continental landmass in its early history (C. Hawkesworth et al. 2019).

Colder simulations conducted outside the OHZ experience atmospheric collapse that prevents them from reaching equilibrium. If these simulations were able to progress to equilibrium without collapsing, the final atmospheric $p\mathrm{CO_2}$ would likely be significantly reduced, altering the cloud formation patterns and resulting spectral signatures of these worlds. A decrease in $p\mathrm{CO_2}$ could also affect cloud distribution and planetary albedo. However, our cold simulations do not reach equilibrium as a result of our fixed $p\mathrm{CO_2}$ values. Therefore, the current cloud maps and spectra derived from these cold simulations might not accurately represent the actual equilibrium state of such planets. We consider the effects of atmospheric collapse when interpreting these results, as they may lead to an overestimation of atmospheric $p\mathrm{CO_2}$. Further studies aimed at stabilizing these cold simulations could provide further constraints on the plausible atmospheric properties of planets beyond the OHZ.

The OHZ boundaries calculated herein were determined considering an anoxic, $CO_2$-dominated atmosphere. The presence of stronger greenhouse gases such as $CH_4$ could also trap more heat, warming the planet and maximizing the greenhouse effect across the infrared spectrum. This would lead to further extension of the OHZ beyond what high $p\mathrm{CO_2}$ alone would allow. However, the climatic impact of $CH_4$ depends strongly on the spectral type of the host star (R. M. Ramirez & L. Kaltenegger 2018). While this study explores a range of $CO_2$ partial pressures, it does not determine whether those $CO_2$ inventories are geochemically attainable for planets orbiting the Sun or TRAPPIST-1. Coupled geochemical and climate modeling will therefore be required to determine which atmospheric $CO_2$ abundances can be sustained over geologic timescales and, consequently, to establish the true extent of the OHZ for a given planetary system. Other geochemical weathering processes beyond the carbonate-silicate cycle may contribute to negative temperature feedbacks, such as seafloor weathering and reverse weathering (T. T. Isson & N. J. Planavsky 2018), but this will not change the maximum greenhouse limit. We intend to explore this in future work.

The opacity tables based on HITRAN 2012 (L. Rothman et al. 2013; M. Turbet et al. 2017a) include $H_2O$ continuum absorption but do not account for the recently characterized $CO_2$ continuum absorption of dense $CO_2$ atmospheres (G. Chaverot et al. 2025). Inclusion of this continuum would increase the infrared opacity of $CO_2$-dominated atmospheres, leading to somewhat stronger greenhouse warming, particularly for the 10–20 bar atmospheres. As a result, the surface temperatures may be slightly underestimated in our high-$CO_2$ atmospheres, but this effect is expected to be limited to several kelvins and would likely not change our overall findings.

### *4.4. Stellar Considerations for Habitability*

Although planets around M dwarfs can be observed with JWST, the habitability of these worlds may be inhibited by high X-ray luminosities (E. L. Shkolnik & T. S. Barman 2014; R. O. P. Loyd et al. 2016; B. A. Binder et al. 2024), extreme water loss in the pre-main-sequence phase (R. Luger & R. Barnes 2015; R. M. Ramirez & L. Kaltenegger 2014; F. Tian 2015), and energetic flares that can lead to atmospheric loss (K. Garcia-Sage et al. 2017; L. N. R. do Amaral et al. 2022). Planets around "Sun-like" stars will likely not be inhibited by such processes, providing a stronger case for the existence of habitable worlds, which will also be more accessible for imaging with the upcoming HWO and ELT. While this work considers the expected OHZ for a G star, we note that K dwarfs may offer even more advantages for habitable planet hosts due to their abundance, extended lifetimes, and increased planet–star contrast ratio that makes them easier to observe in direct imaging (M. Cuntz & E. F. Guinan 2016; G. N. Arney 2019). For stars hotter than a mid-K dwarf ($\sim$4500 K), $CH_4$ enhances greenhouse warming by several tens of degrees, whereas planets around cooler stars exhibit a significant antigreenhouse response (R. M. Ramirez & L. Kaltenegger 2018). While K dwarf aquaplanets near the IHZ are at higher risk of water vapor escape and nightside cold-trapping, the mid HZ and OHZ offer stronger prospects for the detection of water-rich worlds, especially for systems with reduced nightside snow rates (A. H. Lobo & A. L. Shields 2024). We intend to expand this study to include planets around both K and F stars in upcoming work.

## 5. Conclusion

This study investigated the climates and spectral observables of planets in the OHZ with high atmospheric $pCO_2$ using a 3D GCM that includes $CO_2$ condensation. Our results for planets around Sun-like stars show strong (but not necessarily expected) agreement with previous 1D OHZ calculations by R. K. Kopparapu et al. (2014), where both studies find similar maximum greenhouse OHZ limits. However, notable differences arise due to our inclusion of 3D atmospheric circulation with $CO_2$ condensation, which weakens the greenhouse effect and leads to a slightly reduced OHZ boundary of 1.64 au compared to 1.67 au. In contrast, for planets around stars identical to the ultracool dwarf TRAPPIST-1 (M8V), we find a somewhat more favorable maximum greenhouse limit, with a limiting Seff of 0.19 or below for $pCO_2$ = 10–20 bars, compared to 0.22 in R. K. Kopparapu et al. (2013b, 2014). This climate response cannot be captured with 1D models, emphasizing the importance in using 3D GCMs to more accurately determine the limits of the OHZ, especially at high pressures.

Spectral simulations reveal that high $pCO_2$ atmospheres in the OHZ produce unique features in transmission, emission, and reflected light. In cold atmospheres, $CO_2$ surface ice and clouds cause the planet to appear much brighter than temperate worlds, which may allow us to distinguish them from habitable worlds with direct-imaging observations. However, for planets orbiting M dwarf stars such as TRAPPIST-1, transit observations may struggle to identify between spectral features on cold, nonhabitable or planets with limited fractional habitability due to the presence of a high-altitude $CO_2$ cloud deck. These clouds raise the continuum level and make the atmosphere appear more opaque at most wavelengths, muting the absorption features from $H_2O$ and $CO_2$. This could make it difficult to distinguish between a thick or thin atmosphere. On the other hand, warm planets with temperate regions near the terminator can possess deep atmospheric features with limited $CO_2$ clouds. Future transit observations should be complemented by other assets, such as high-resolution ground-based observations, to overcome these limitations and better characterize the atmospheres of planets around M dwarfs in the OHZ. Direct-imaging thermal infrared observations of OHZ planets are not yet well explored, but we find that significant differences between habitable and nonhabitable $CO_2$-dominated worlds may be an important objective for follow-up work.

## Acknowledgments

D.P. gratefully acknowledges support from the NASA FINESST Fellowship program issued via grant No. 80NSSC22K1319. Computations were performed using the computer clusters and data storage resources of the HPCC, which were funded by grants from NSF (MRI-2215705, MRI-1429826) and NIH (1S10OD016290-01A1). This work was further supported by the NASA Interdisciplinary Consortia for Astrobiology Research (ICAR) program via the CHAMPs (Consortium on Habitability and Atmospheres of M dwarf Planets) team with funding issued under grant No. 80NSSC23K1399 and the Alternative Earths team with funding issued under grant No. 80NSSC21K0594. T.J.F. acknowledges support from the GSFC Sellers Exoplanet Environments Collaboration (SEEC), which is funded by the NASA Planetary Science Divisions Internal Scientist Funding Model. M.T. acknowledges support from the Tremplin 2022 program of the Faculty of Science and Engineering of Sorbonne University. M.T. acknowledges support from BELSPO BRAIN (B2/212/PI/PORTAL). We thank Ravi Kopparapu for helpful input regarding our model comparison.

## ORCID iDs

Daria Pidhorodetska https://orcid.org/0000-0001-9771-7953
Edward W. Schwieterman https://orcid.org/0000-0002-2949-2163
Thomas J. Fauchez https://orcid.org/0000-0002-5967-9631
Martin Turbet https://orcid.org/0000-0003-2260-9856